\documentclass[10p]{article}

\usepackage[a4paper, left=4cm, right=4cm, top=5.7cm, bottom=5.1cm, headsep=0.8cm]{geometry}

\usepackage{mathptmx}      
\newcommand{\mathsym}[1]{{}}
\newcommand{\unicode}[1]{{}}

\newcommand{\RR}{{\cal R} }
\newcommand{\WW}{{\cal W} }
\newcommand{\MM}{{\cal M} }

\newcommand{\tWeylC}{C}
\newcommand{\tWeylCdual}{{ }^*\!C}
\newcommand{\cm}{Car\-mi\-na\-ti--McLena\-ghan }
\newcommand{\sch}{Schwarzschild }
\newcommand{\cc}{{\it ccgrg} }
\newcommand{\mat}{{\it Mathematica} }
\newcommand{\ddager}{{} }
\begin{document}

\renewcommand{\thefootnote}{\fnsymbol{footnote}}

\normalsize

\centerline{ANDRZEJ WOSZCZYNA\footnote{Institute of Physics, Cracow University of Technology; Copernicus Center for Interdisciplinary Studies; uowoszcz@cyf-kr.edu.pl}, PIOTR PLASZCZYK\footnote{Astronomical Observatory, Jagiellonian University;  piotr.plaszczyk@uj.edu.pl},}
\centerline{ WOJCIECH CZAJA\footnote{Copernicus Center for Interdisciplinary Studies; wojciech.czaja@gmail.com}, ZDZIS{\L}AW   A. GOLDA\footnote{Astronomical Observatory, Jagiellonian University; Copernicus Center for Interdisciplinary Studies;\\ zdzislaw.golda@uj.edu.pl}}

\bigskip\bigskip 

\normalsize 

\begin{center}
\Large SYMBOLIC TENSOR CALCULUS --- FUNCTIONAL  AND~DYNAMIC APPROACH
\end{center}

\hrule

\normalsize

\leftline{A\,b\,s\,t\,r\,a\,c\,t}

\begin{quote}
\footnotesize 
In this paper, we briefly discuss the dynamic and functional approach to computer symbolic tensor analysis. The {\em ccgrg\/} package  for Wolfram Language/Mathematica is used to illustrate this approach. Some examples of applications are attached.

\end{quote} 

\normalsize

\section{Introduction}
\label{intro}

Cartesian tensors are identified with matrices.
In contrast,  in curved coordinates, even more so in a curved space, viewing  the tensor matrix {\it in extenso\/} is not instructive. The subject of interpretation are equations, invariants, symmetries, and observables, while the tensor components values ---  except for certain privileged frames of reference --- are of minor importance. This follows directly from the general principle of covariance, which states that the physical sense is independent of the choice of coordinate system, whereby each coordinate choice is allowed. 

The vast majority of the computer tensor tools --- including those build in Wolfram Language/Mathematica, the symbolic language which we use here --- identify tensors with matrices, providing them with procedures for tensor contraction, tensor product and the rules of symmetry. Objects of such a class require the evaluation of complete tensor matrices. A sequence of intermediate steps evaluates quantities, most of which are irrelevant to the actual goal of the calculation, and are an unnecessary burden. A few tensor packages can operate on abstract level, where the differential manifold is not specified. They cover the entire tensor algebra, and essentially, form the sophisticated programming languages \cite{MathTensor,grtensor,xAct}. On the other hand, all of them depend on the primary computer algebra languages they employ (Mathematica, Mapple) --- the languages which  constantly develop. The risk of losing compatibility between the primary language and the computer algebra package is greater the more complex the package.

The project we realized in 2013--2014 was aimed at constructing an open source package~\cite{ccgrg02} for Mathematica, which defines some basic concepts and rules, and is open to development by users. The syntax is kept as close as possible to standard textbook notation \cite{Synge} and typical scheme of manual calculations. On the other hand, we employ dynamic programming with modern computational techniques, a {\it call-by-need strategy\/} and {\it memoization}. The question is not how much can be computed, but how flexible the software can be? The results are discussed in this paper.

We have taken into account that physicists have in mind co- and contravariant tensor components, rather than  individual tensors of a precised valence (covariant or contravariant tensors). Rising and lowering tensor indices is a basic manual technique which comes directly from the Einstein convention. What physicist actually do is they aggregate tensors of arbitrary valence (all combinations of upper and lower indices) into a single object. The choice of particular valence is left for the future, and decided upon later.
In order to stick to that, while the matrix representation is employed, one would need to evaluate and store all the co-  and contravariant index combinations. In the case of the Riemann tensor in four dimensions, this gives $8^4$ expressions, while only a small proportion of them will probably be used further on. Although they are not independent as subjects to tensor symmetries, the benefits of this fact are not straightforward. Modern techniques of differentiation, and reducing terms are quick enough, while the evaluation of conditions {\em If} still cost time. As a result, computing the same component twice may cost only slightly more than retrieving it from another matrix element. But there still remains the problem of memory. In the matrix approach, thousands of array's components are build of long algebraic expressions. 

The functional representation of tensors allows one to avoid some of these problems. In this paper, we focus on the {\it ccgrg}-package (the open source code written for {\it Mathematica 9--10\/}, and distributed from Wolfram Library~\cite{ccgrg01}). The goal of the article is to argue for some particular computation technique. We do not give here the precise tutorial to the package. The same approach to tensors in Python is presented in~\cite{GraviPy}.

\section{Dynamic paradigm of programming}

The idea is not new, as it dates back to the famous Richard Bellman book {\em Dynamic Programming~(1957)}. Bellman introduces the concept of  the goal function. This means that the starting point is identical to the final expression in the search. To evaluate this expression, the algorithm automatically splits tasks into sub tasks down to elementary level. The tree of calls is not explicitly known to programmers, and the computational process does not require any intelligent interference.
For tensors, this means that the subroutines do not evaluate all the components but some of them, acting on demands of routines from a higher level. Most tensor components remains unknown. Only these are computed which contribute to the final expression. This process is known as {\em lazy evaluation\/} or {\em call-by-need strategy\/}~\cite{lazy,Reynolds}. 
To evaluate Ricci scalar for~\sch metrics (\ref{emptystatic}),~{\em ccgrg\/}
calls $16$~Riemann components. $40$~Riemann components are needed for full Ricci tensor, and $808$~Riemann components for 
\cm $\WW_1$~invariant~\cite{CM} 
$
\WW_1=\frac{1}{8}
 (
 \tWeylC_{a}{}_{b}{}_{c}{}_{d}{}+
{\tt i}\,  \tWeylCdual_{a}{}_{b}{}_{c}{}_{d}{} 
) \tWeylC^{a}{}^{b}{}^{c}{}^{d}{}
$ 
build of the Weyl $\tWeylC$ and dual Weyl $\tWeylCdual$ contractions. In each case, the appropriate components are selected automatically and the rest of components remains unevaluated. 

Tensors form a class of functions, where the tensor indices are the innermost group of arguments. For instance, the covariant derivative we denote as:
\begin{eqnarray}
	 T_{ij;m} &\rightarrow& \tt \nabla[T][{i,j},{m}],\\
	 T_{ijk;mn} &\rightarrow& \tt \nabla[T][{i,j,k},{m,n}].
\end{eqnarray}
The outer argument contains the tensor name, while the inner argument lists the tensor indices.
In the case of Lie derivative
\begin{eqnarray}
	\pounds_{U}V_{i} &\rightarrow & \tt LieD[U][V][i],\\
	\pounds_{U}T_{ij} &\rightarrow & \tt LieD[U][T][i,j],
\end{eqnarray}
the outermost argument $U$ is the vector field that generates the dragging flow, the middle ($\tt V$ or $\tt T$) is the tensor name, and the innermost, according to the general rule, specify indices. 
The grouping of arguments, and the appropriate ordering of groups allows one to work with headers $\tt \nabla[T]$, $\tt LieD[U][T]$, etc., treating them as operators. Inserting new groups of outer arguments does nor affect  the methods governing tensor index operations. Indexes run through $\{1,\ldots,\makebox{dim}\}$ for covariant components and through $\{-\makebox{dim},\ldots,-1\}$ for contravariant components (the inverse Parker--Christensen notation~\cite{MathTensor}). Thus we write:
\begin{eqnarray}
	 T_{ij} &\rightarrow& \tt T[i,j],\\
	 T_{i}^{j} &\rightarrow& \tt T[i,-j],\\
	 T^{ij} &\rightarrow& \tt T[-i,-j],\\
	 T^{ij}{}{}_{k} &\rightarrow& \tt T[-i,-j,k],
\end{eqnarray}
and so on. Tensor valence operations (conversions from covariant to contravariamt components, and vice versa), and the covariant differentiation are the only methods of the object. Other tensor operations like contractions, products, etc.~are realized by elementary algebra and the Einstein convention. Conversions are called just by placing or removing the sign ``minus'' in front of selected tensor indices.

\section{The manifold}

Tensors live on a manifold. 
This manifold is declared in the opening procedure {\tt open[x,g]}, where the list of coordinate names $x$ and metric tensor $g$ are the arguments. When the manifold is `opened', tensors are evaluated on this particular manifold. The metric tensor can be uniquely defined, or may contain arbitrary functions of coordinates. In the last case, ${\tt open}$ indicates the class of manifolds distinguished by the specified structure of the metric tensor.

Tensors are created by setting object data in the form of the pair: tensor matrix, and the associated tensor valence (all covariant components as a default). By calling\break {\tt tensorExt[matrix, valence]} one creates a tensor object which can return components for arbitrary valence, on demand. Tensors which are in frequent use (Ricci, Riamann, Weyl, the first and the second fundamental forms on hypersurfaces), as well as covariant and Lie derivatives are predefined in the {\em ccgrg\/}-package. 
Tensor components are not evaluated in advance, only when called for the first time. This is guaranteed by the SetDelayed $(:=)$ command. For instance, when calling the Riemann curvature tensor in the \sch spacetime one obtains:
 	\begin{eqnarray}
R_{1212}\rightarrow \tt In[1]&:=&\tt tRiemann[1,2,1,2]\nonumber\\
\tt Out[1]&:=& -\frac{2M}{r^3} \\
R_{1}{}_{2}{}_{1}{}^{2}{}\rightarrow \tt In[2]&:=&\tt tRiemann[1,2,1,-2]\nonumber\\
\tt Out[2]&:=& -\frac{2M(2M-r)}{r^4}\\ 
R_{ijmn}R^{ijmn}\rightarrow \tt In[3]&:=&\tt Sum[tRiemannR[i, j, m, n] tRiemannR[-i, -j, -m, -n] \nonumber\\
& & \tt \{i, dim\}, \{j, dim\}, \{m, dim\}, \{n, dim\}]\nonumber\\
\tt Out[3]&:=& 48\frac{M^2}{r^6} 
	\end{eqnarray}
The tensor rank is fixed, except for the covariant derivative, where each differentiation rises the rank by one. 
Covariant differentiation is realized by header $\tt \nabla$ (or its full name, {\tt cov\-ariantD}), and by extending the list of indices by one. Extension of this list by more than one returns derivatives of a higher order. 
	\begin{eqnarray}
R_{1}{}_{2}{}_{1}{}^{2}{}_{;2}{}\rightarrow \tt In[1]&:=&\tt \nabla[tRiemannR][1, 2, 1, -2, 2]\nonumber\\
\tt Out[1]&:=& -\frac{6M(2M-r)}{r^5} \\
R_{1}{}_{2}{}_{1}{}^{2}{}_{;2}{}_{2}{}\rightarrow \tt In[2]&:=&\tt \nabla[tRiemannR][1, 2, 1, -2, 2, 2]\nonumber\\
\tt Out[2]&:=& -\frac{6M(9M^2-4M r)}{r^6}
	\end{eqnarray}
Rising the tensor rank by differentiation does not affect the general ability to rise or lower indices by choosing appropriate signs, both for the original and the differentiation indices. 
The flexibility of tensor indexes stems directly from the functional representation of tensor and  the {\em call-by-need strategy\/}.

\section{Memoization and symmetries}

Trees of calls for different tasks may have nonempty intersections (and typically this is the case). The {\tt SetDelayed} function which enables {\em call-by-need strategy\/} at the same time forces the multiple evaluation of components whenever trees of  calls overlap. To avoid unwilling effects, one needs to store computed values, and to have a mechanism that can recognize known and unknown components. This process is called {\em memoization}. The term {\it memoization} was introduced by Donald Michie in 1968~\cite{memo} and refers to the class of functions that can learn what they found in past calls. In $C$ or Python, the {\it memoization} tools are usually constructed by programmers (see~\cite{GraviPy}). Wolfram Mathematica provides dedicated tool in the core language. The construction consists in the recursive definition of the form
	\begin{equation}
f[x\_]:=f[x]=expr[x,\ldots]
	\end{equation}
where the same name of the function appears twice, first followed by {\tt SetDelayed} and next, by {\tt Set}. The definition of $f$ which is initially equivalent to $expr[x,\ldots]$, is successfully supplemented by the values found during each function call. The memoized (memorized) values form cache. The cache is searched first; therefore, no tensor component is evaluated twice. This means that each of the expressions in the examples (10)--(14) is evaluated only once. Whenever called again, the  values are instantly retrieved from the cache. The same refers to expressions called by subroutines. In the \sch space time (\ref{emptystatic}),  the \cm invariants $\WW_1$ and $\WW_2$, when evaluated separately, call 808 and 552 Riemann components, respectively. Jointly, they call 1046 Riemann components which means that 296 components are taken from the cache. 

Calls appeal to tensor symmetries. The symmetry rules are the argument conditions within the same memoizing function scheme
\begin{equation}
{\tt f[x\_]/;}~\langle\makebox{test[x]}\rangle:={\tt f[x]}=expr[x,\ldots]
\end{equation} 
For instance, for the Ricci tensor $R_{ij}$, the conditional definition takes the form
\begin{eqnarray}
& & \tt tRicciR[i\_, j\_] /; i > j := \tt tRicciR[i, j] = tRicciR[j, i]\\
& & \tt tRicciR[i\_, j\_] := \tt tRicciR[i, j] =Sum[tRiemannR[-s,i,s,j],\{s, dim\}]\nonumber
\end{eqnarray}
with positive $i$, $j$ (covariant components).  The rules of symmetry are introduced directly in the definition of the tensor. In order to recognize Ricci symmetries, the kernel does not refer to the symmetries of the Riemann tensor. Subsequently, the Riemann tensor symmetries are not the result of the evaluation of Christoffel symbols, etc. Symmetries defined as the argument conditions in lazy-evaluated functions put constraints directly on computational processes.

\section{The confidence of results in dynamic programming}

Dynamic programming allows one to reach the goal at a minimal cost of time and memory. The {\it call-by-need strategy\/} avoids computing unnecessary components. Memoization protects against multiple evaluations.  The cost we pay, however, is confidence in the  results. Functions that can learn and remember are reluctant to forget. The user must not change the tensor definition during the same session. The cache has priority over evaluation. A new definition does not affect the values which are already found. If definitions change, one may readily collect a mixture of results belonging to different definitions.

The content of the cache together with the general tensor definition can be viewed by the Mathematica command ${\tt Definition[tensorname]}$.
Based on this command,\break ${\tt cacheview[tensorname]}$ returns the list of the {\it memorized} tensor components. For instance
	\begin{eqnarray}
\tt In[26]:=& & \tt cacheview[tRiemannR]\nonumber\\
\tt Out[26]:=& & \tt \{\{tRiemannR, \{-4, 1, 4, 1\}\}, \{tRiemannR, \{-3, 1, 3, 1\}\}\}\label{sam}
	\end{eqnarray}
means that $R^{4}{}_{1}{}_{4}{}_{1}{}$ and $R^{3}{}_{1}{}_{3}{}_{1}{}$ Riemann components have been already called and stored in the cache. To view all the memoized quantities which refer to $\tt tRiemannR$, one writes
	\begin{eqnarray}
\tt In[27]:=& & \tt associated[tRiemannR]\nonumber\\
\tt Out[27]:=& & \tt \{\{tRiemannR, \{-4, 1, 4, 1\}\}, \{tRiemannR, \{-3, 1, 3, 1\}\}\nonumber\\
& & \tt\{covariantD[tRiemannR], \{1, 2, 1, 2, 2\}\}\}\}\label{razem}
	\end{eqnarray}
Commands
	\begin{eqnarray}
\tt In[31]:=& & \tt retreat[tRiemannR]\nonumber\\
\tt In[32]:=& & \tt retreat[tRiemannR, associated]
	\end{eqnarray}
clear the stored values displayed in (\ref{sam}) and (\ref{razem}), respectively. In contrast to the core-languge  $\mat$ commands {\tt Clear} and {\tt Unset}, thr command {\tt retreat} does not affect the general definition of a tensor. Calling {\tt cacheview} after {\tt retreat} returns the empty list.

For sake of safety, the name of each memoizind tensor shall be appended to\break  ${\tt memRegistry}$. This is realized by {\tt erasable[tensorname]}. ${\tt memRegistry}$ allows the automatic erasing of the whole cache content whenever ${\tt open[x,g]}$ is called. User may safely redefine the manifold with no risk of unwanted remnants from the past computations. $\cc$ is equipped with extensive random tests of memory clearance. The above mentioned security tools do not prevent some incidental mistakes such as those shown below 
	\begin{eqnarray}
\tt In[41]:=& & \tt f[k\_]:=f[k]=k;f/@Range[8];\nonumber\\
\tt In[42]:=& & \tt f[k\_]:=f[k]=1/k;f/@Range[8];\nonumber
	\end{eqnarray}
To effectively implement {\it memoization\/} to algorithms, minimal training and discipline are indispensable.

\section{Tensor definition scheme}

This is the place to show the typical construction of a tensor. We choose the induced metrics (the first fundamental form) $h_{ij}=g_{ij}-v_i v_j$ on the hypersurface orthogonal to the vector field $v_i$ in the space with the metric $g_{ij}$. 
	\begin{eqnarray}
\tt & & \tt erasable/@\{hcov,h\ddager \};\label{tensordef}\\
\tt & & \tt hcov[v\_][j\_,i\_]/;i<j:=hcov[v][j,i]=hcov[v][i,j];\nonumber\\
\tt & & \tt hcov[v\_][i\_,j\_]:=g\ddager [i,j]-v[i]v[j]/vectorsquared[v]//simp[];\nonumber\\
\tt & & \tt h\ddager [v\_][i\_,j\_]/;indeX[i,j]:=h\ddager[v][i,j]=tensorExt[hcov[v]][i,j]//simp[];\nonumber
	\end{eqnarray}
In the first line we append ${\tt memRegistry}$. The next line contains the symmetry settings. The third line defines the matrix of the covariant tensor ({\tt hcov} is not a tensor in the meaning of the  $\cc$ package and will be invisible in the general context.) The last line creates the tensor {\tt h} as an object containing data ({\tt hcov}), methods to control the index values (condition {\tt indeX}), and methods to rise or lower indices (provided by the {\tt tensorExt} procedure). 

Following the scheme above, users may define their own tensors and append to the package, or just use them in notebooks. Summation, multiplying tensors by numbers or functions (not containing indices as arguments!), and covariant differentiation return tensors which not need to be separately defined as in (\ref{tensordef}). Yet, even in these cases, the definition scheme (\ref{tensordef}) is strongly recommended to assure  effectiveness.

\section{Examples of use}

\subsection{Differential operators in an arbitrary coordinate system}

In technical sciences there is a need to express differential operators in some particular curved coordinates or on curved surfaces. The task may appear as a part of the heat transport or diffusion problems for systems of more complex geometry. Many typical gradient or Laplacian expressions are catalogued in the literature, but still the range of geometries that can be encountered in nature is much richer. Below, we choose a catenoid --- one of the best known minimal surfaces --- for which, we hope, Laplacian is not published in print.  

The family of catenoids numbered by $r$ and parametrized by $u$ and $v$ define a curved coordinate system in the Euclidean space. The first step is to find the Euclidean metric form in coordinates $r, u, v$ (the core Wolfram Language)
	\begin{eqnarray}
\tt In[1]:=& & \tt Needs["ccgrg`"];\nonumber\\
\tt In[2]:=& & \tt x[r\_,u\_, v\_] = r Cosh[v/r] Cos[u];\nonumber\\
\tt In[3]:=& & \tt y[r\_,u\_, v\_] = r Cosh[v/r] Sin[u];\nonumber\\
\tt In[4]:=& & \tt z[r\_,u\_, v\_] = v;\nonumber\\
\tt In[5]:=& & \tt crd = \{r, u, v\};\nonumber\\
\tt In[6]:=& & \tt form = Dt[x[r,u, v]]^2 + Dt[y[r,u, v]]^2 + Dt[z[r,u, v]]^2 // FullSimplify\nonumber\\\nonumber\\
\tt Out[6]:=& & \tt Cosh[\frac{v}{r}]^2 (r^2 Dt[u]^2 + Dt[v]^2) + 
\tt \frac{Dt[r]^2 (r Cosh[\frac{v}{r}] - v Sinh[\frac{v}{r}])^2}{r^2} +\nonumber \\
\tt & &\tt \frac{Dt[r] Dt[v] (v - v Cosh[\frac{2v}{r}] + r Sinh[\frac{2v}{r}])}{r}\label{forma}
	\end{eqnarray}
The second step is to evaluate the covariant differentiation $f_{,i}{}^{;i}{}$ of an arbitrary function f on the manifold with the metric form (\ref{forma}):
	\begin{eqnarray}
\tt In[7]:=& & \tt open[crd, toMatrix[form, crd]];\nonumber\\
\tt In[8]:=& & \tt Sum[\nabla[f[r, u, v]][i, -i], \{i, dim\}] // FullSimplify\nonumber\\\nonumber\\
\tt Out[8]:=& & 
\tt \frac{r^2 f^{(2,0,0)}[r,u,v]}{\left(r-v Tanh\left[\frac{v}{r}\right]\right)^2}
+\frac{{Sech}^2\left[\frac{v}{r}\right] f^{(0,2,0)}[r,u,v]}{r^2}+f^{(0,0,2)}[r,u,v]+\nonumber\\
& &\tt \frac{2 r f^{(1,0,1)}[r,u,v]}{v-r Coth\left[\frac{v}{r}\right]}
\tt -\frac{\left(r Sinh \left[\frac{2 v}{r}\right]+2 v\right)^2{Sech}^4\left[\frac{v}{r}\right] f^{(1,0,0)}[r,u,v]}{4 \left(r-v Tanh\left[\frac{v}{r}\right]\right)^3}\nonumber\\
	\end{eqnarray}
The only time-consuming operations are simplifications $\tt FullSimplify$, which are inevitable in this code. Similar computations can be easily performed for an arbitrary transformation defined by inputs the $\tt In[2]{-}In[4]$.

\subsection{Riemann curvature invariants: the case of spherical symmetry}

\cm invariants~\cite{CM} form a complete set of Riemann invariants for the spacetime which obey the Einstein equations with the hydrodynamic energy-momentum tensor. These invariants read
	\begin{eqnarray}
\RR_1&=&\frac{1}{4} S^{a}{}_{b}{} S^{b}{}_{a}{},\\
\RR_2&=&-\frac{1}{8} S^{a}{}_{b}{} S^{b}{}_{c}{} S^{c}{}_{a}{},\\
\RR_3&=&\frac{1}{16} S^{a}{}_{b}{} S^{b}{}_{c}{} S^{c}{}_{d}{} S^{d}{}_{a}{}\label{R123},\\
\WW_1&=&\frac{1}{8}
 (
 \tWeylC_{a}{}_{b}{}_{c}{}_{d}{}+
i  \tWeylCdual_{a}{}_{b}{}_{c}{}_{d}{} 
) \tWeylC^{a}{}^{b}{}^{c}{}^{d}{},
\\
\WW_2&=&
-\frac{1}{16} (
 \tWeylC_{a}{}_{b}{}^{c}{}^{d}{}+
i  \tWeylCdual_{e}{}_{f}{}^{a}{}^{b}{}{}
)  \tWeylC_{c}{}_{d}{}^{e}{}^{f}{}  \tWeylC_{e}{}_{f}{}^{a}{}^{b}{}\label{W12},\\
\MM_1&=&\frac{1}{8} S^{a}{}^{d}{} S^{b}{}^{c}{} (- i\tWeylCdual_{a}{}_{b}{}_{c}{}_{d}{} + \tWeylC_{a}{}_{b}{}_{c}{}_{d}{}),\\
\MM_2&=&\frac{1}{8} i\tWeylCdual_{a}{}_{b}{}_{c}{}_{d}{} S^{b}{}^{c}{} S_{e}{}_{f}{} \tWeylC^{a}{}^{e}{}^{f}{}^{d}{} + \frac{1}{16} S^{b}{}^{c}{} S_{e}{}_{f}{} (\tWeylC_{a}{}_{b}{}_{c}{}_{d}{} \tWeylC^{a}{}^{e}{}^{f}{}^{d}{} - \tWeylCdual_{a}{}_{b}{}_{c}{}_{d}{} \tWeylCdual^{a}{}^{e}{}^{f}{}^{d}{}),\\
\MM_3&=&\frac{1}{16} S^{b}{}^{c}{} S_{e}{}_{f}{} \tWeylC_{a}{}_{b}{}_{c}{}_{d}{} \tWeylC^{a}{}^{e}{}^{f}{}^{d}{} + \frac{1}{16} S^{b}{}^{c}{} S_{e}{}_{f}{} \tWeylCdual_{a}{}_{b}{}_{c}{}_{d}{} \tWeylCdual^{a}{}^{e}{}^{f}{}^{d}{},\\
\MM_4&=&-\frac{1}{32} S^{a}{}^{g}{} S^{c}{}_{d}{} S^{e}{}^{f}{} \tWeylC_{a}{}_{c}{}^{d}{}^{b}{} \tWeylC_{b}{}_{e}{}_{f}{}_{g}{} -\frac{1}{32} S^{a}{}^{g}{} S^{c}{}_{d}{} S^{e}{}^{f}{} \tWeylCdual_{a}{}_{c}{}^{d}{}^{b}{} \tWeylCdual_{b}{}_{e}{}_{f}{}_{g}{},\\
\MM_5&=&\frac{1}{32} S^{b}{}^{c}{} S^{e}{}^{f}{} (i\tWeylCdual^{a}{}^{g}{}^{h}{}^{d}{} + \tWeylC^{a}{}^{g}{}^{h}{}^{d}{}) (\tWeylCdual_{a}{}_{b}{}_{c}{}_{d}{} \tWeylCdual_{g}{}_{e}{}_{f}{}_{h}{} + \tWeylC_{a}{}_{b}{}_{c}{}_{d}{} \tWeylC_{g}{}_{e}{}_{f}{}_{h}{}).
\label{mixed}
	\end{eqnarray}
$S_{ab}$ stands for the Plebanski tensor (the traceless Ricci tensor $S_{ab}=R_{ab}-\frac{1}{4}R g_{ab}$). Invariants $\WW_1$ and $\WW_2$ are complexes, which are defined  by means  of contractions of the Weyl $ \tWeylC$  and the dual Weyl  $ \tWeylCdual$ tensors.

In the case of vacuum and spherically symmetric spacetime (Black Hole)
	\begin{equation}
ds^2=
-\left(1-\frac{2M}{r}\right) dt^2
+\left(1-\frac{2M}{r}\right)^{\!-1}dr^2
+r^2 \left( d\vartheta^2+\sin^2\vartheta d\varphi^2\right),
\label{emptystatic}
	\end{equation}
the complete computation appears as follows:
	\begin{eqnarray}
\tt In[1]:=& & \tt Needs["ccgrg`"];\nonumber\\
\tt In[2]:=& & \tt x = \tt\{t, r, \vartheta, \varphi\};\nonumber\\
\tt In[3]:=& & \tt g :=\tt DiagonalMatrix[{-(1-2M/r), (1-2M/r)^{-1}, r^2, r^2 Sin[\vartheta]]^2}];\nonumber\\
\tt In[4]:=& & \tt \$Assumptions = \tt And @@\{0 < t , 0 < r , 0 < \vartheta < \pi , 0 < \varphi < 2 \pi , 0 < M \};\nonumber\\
\tt In[5]:=& & \tt open[x, g];\nonumber\\
\tt In[6]:=& & \tt \{CMinvR1,CMinvR2,CMinvR3\}\nonumber\\
\tt Out[6]:=& & \tt \{0,0,0\}\nonumber\\
\tt In[7]:=& & \tt \{CMinvW1,CMinvW2\}\nonumber\\
\tt Out[7]:=& & \tt \{6M^2/r^6,-6M^2/r^6\}\nonumber\\
\tt In[8]:=& & \tt \{CMinvM1,CMinvM2,CMinvM3,CMinvM4,CMinvM5\}\nonumber\\
\tt Out[8]:=& & \tt \{0,0,0,0,0\}\nonumber\\
\tt In[9]:=& & \tt \{TimeUsed[], \$ MachineType, \$ Version, \$ ProcessorType, \nonumber\\
& &\tt \$ ProcessorCount\} \nonumber\\
\tt Out[9]:=& & \tt \{8.80264, PC, 10.0 for Linux x86 (64\mbox{-}bit) (December 4, 2014), x86\mbox{-}64, 2\}\nonumber\\
	\end{eqnarray}
As see result of these operations, two nonvanishing invariants were found, while 1320 of 4096 components of the Riamann tensor object were evaluated and {\tt memorized}.
For the PC computer with AMD64 processor and Linux Mint x86 (64-bit), the whole operation took 8.80264 seconds CPU.

\section{Final remarks}
The degree of complexity in a symbolic programming is much less predictable than in numerical computation. This is due to the fact that the quantity of algebraic terms in an expression is not a uniquely determined number. This number significantly depends on the ability of the algorithm to recognize mathematical identities, and for the same function, may differ in orders of magnitude. In this circumstance, the economic style of the dynamic programming  ({\it call-by-need strategy, memoization}) may take the first rank role.

\bigskip

\noindent
{\it The research supported  by grant from the John Templeton Foundation.}

\end{document}